\documentclass[english,prc,superscriptaddress,showpacs,preprintnumbers,amsmath,amssymb,floatfix]{revtex4}
\usepackage[T1]{fontenc}
\usepackage[latin9]{inputenc}
\usepackage{textcomp}
\usepackage{amstext}

\makeatletter

\newcommand{\lyxmathsym}[1]{\ifmmode\begingroup\def\b@ld{bold}
  \text{\ifx\math@version\b@ld\bfseries\fi#1}\endgroup\else#1\fi}

\@ifundefined{textcolor}{}
{%
 \definecolor{BLACK}{gray}{0}
 \definecolor{WHITE}{gray}{1}
 \definecolor{RED}{rgb}{1,0,0}
 \definecolor{GREEN}{rgb}{0,1,0}
 \definecolor{BLUE}{rgb}{0,0,1}
 \definecolor{CYAN}{cmyk}{1,0,0,0}
 \definecolor{MAGENTA}{cmyk}{0,1,0,0}
 \definecolor{YELLOW}{cmyk}{0,0,1,0}
 }

\usepackage{dcolumn}\usepackage{bm}\usepackage{longtable}

\makeatother

\usepackage{babel}
\begin{document}

\title{Zeros of 6-j symbols and more:Atoms,Nuclei and Bosons.}

\author{L. Zamick}

\affiliation{Department of Physics and Astronomy, Rutgers University, Piscataway
NJ 08854 }

\author{S. J.Q. Robinson}

\affiliation{Department of Physics, Millsaps College, Jackson, MS 39210, }

\date{\today}
\begin{abstract}
The absence of certain LS states in atoms leads to the vanishing of
several 6-j symbols.One of these vanishing 6-j's explains the absence
of a certain jj coupling state in a nucleus while the other explains
the vanishing of a certain state for a system of three bosons.This
is part of a continuing study of {}``companion problems''. It is
noted that the vanishing 6-j's play an important role for establishing
partial dynamical symmetries. Whenever possible we offer alternate
explanations that do not involve 6-j symbols. Extensions to vanishing
9-j symbols are also shown. Regge symmetries help to make connections
between different topics. 
\end{abstract}
\maketitle

\section{Introduction}

In this work we address the problem of missing states for certain
configurations of fermions-both in LS coupling and jj coupling, and
also missing states for bosons. Unity is brought to these a'priori
different topics by relations involving 6-j symbols which have similar
appearances in the above three categories. Many, but not all, of the
relations involve the vanishings of 6-j symbols. Extensions to 9-j
symbols are also shown as well as applications to partial dynamical
symmetries. Some results presented here are familiar, some are not.
The main virtue of this work is to bring these diverse topics all
in one place. The topics addressed are femions in LS coupling, fermions
in jj coupling and spinless bosons.

\section{LS Coupling in Atoms and Nuclei}

In a 1989 paper by Judd and Li \cite{judd89} it was noted that for
three electrons in the g$^{3}$ configuration (LS coupling) there
is no quartet state $^{4}$D. For those not familiar with the atomic
notation the superscript refers to the spin degeneracy. In the context
of nuclear physics we would say that the spin S is equal to 3/2. They
were able to show that the non-existence of the quartet state was
due to the vanishing of two 6-j symbols. 
\begin{eqnarray}
\left\{ \begin{array}{ccc}
4 & 2 & 5\\
4 & 4 & 3
\end{array}\right\} =0\qquad\left\{ \begin{array}{ccc}
4 & 8 & 7\\
4 & 4 & 5
\end{array}\right\} =0
\end{eqnarray}

This was generalized to the following relations for even L.

\begin{eqnarray}
\left\{ \begin{array}{ccc}
L & 2 & (L+1)\\
L & L & (L-1)
\end{array}\right\} =0\\
\left\{ \begin{array}{ccc}
L & (3L-4) & (2L-1)\\
L & L & (2L-3)
\end{array}\right\} =0
\end{eqnarray}

We will consider not only 3 g electrons but 3 L electrons where L
is even. We take as given that there are no states of this configuration
with S=$\frac{3}{2}$ and $L_{T}=2$, the latter being the total angular
momentum, i.e. there are no $^{4}$D states of the L$^{3}$ configuration
for any even L. We then find out what are the mathematical consequences.
This will involve relations among the 6-j symbols.The results will
of course also apply to three identical nucleons in LS coupling -3
neutrons or 3 protons. In the nuclear case LS coupling is a better
approximation for light nuclei whereas jj coupling is better for heavier
nuclei.

Note that the S=$\frac{3}{2}$ is the maximum spin for three electrons.
The S=$\frac{3}{2}$ spin wave function must be symmetric since the
$M_{S}=\frac{3}{2}$ state must have all three electrons with spin
up. Hence the orbital part of the wave function must be antisymmetric.

First we couple 2 of the electrons to a godparent angular momentum
L$_{G}$ which must be odd so that the two electrons have an antisymmetric
wave function. The possible values of L$_{G}$ are L-1 and L+1 and
there is no loss in generality in choosing the former.

We then antisymmetrize the state {[}{[}LL{]}$^{L-1}$ L{]}$^{2}$.
\begin{equation}
\Psi={[}1-2(2L-1)\left\{ \begin{array}{ccc}
L & L & (L-1)\\
L & 2 & (L-1)
\end{array}\right\} {]}{[}{[}LL{]}^{L-1}L{]}^{2}-2\sqrt{(2L-1)(2L+3)}\left\{ \begin{array}{ccc}
L & L & (L-1)\\
L & 2 & (L+1)
\end{array}\right\} {[}{[}LL{]}^{L+1}L{]}^{2}
\end{equation}

Since $\Psi$ is zero, the coefficients of the 2 basic states in Eq.
(4) must vanish. This leads to Eq. (2) for the second term while for
the first term we obtain 
\begin{equation}
1-2(2L-1)\left\{ \begin{array}{ccc}
L & L & (L-1)\\
L & 2 & (L-1)
\end{array}\right\} =0
\end{equation}

These relations can be verified case by case from tables of 6-j symbols.

We next show that the non-existence of states with S=$\frac{3}{2}$
$L_{T}=(3L-4)$, also of the L$^{3}$ configuration with even L, leads
to other relations involving 6-j symbols. We choose the value of the
godparent as $L_{G}=(2L-3)$. Antisymmetrizing $\Psi$ = A {[}{[}LL{]}$^{2L-3}$
L{]}$^{3L-4}$ + B {[}{[}LL{]}$^{2L-1}$ L{]}$^{3L-4}$ gives us

\begin{equation}
A=1-2(4L-5)\left\{ \begin{array}{ccc}
L & L & (2L-3)\\
L & (3L-4) & (2L-3)
\end{array}\right\} 
\end{equation}

\begin{equation}
B=2\sqrt{(4L-5)(4L-1)}\left\{ \begin{array}{ccc}
L & L & (2L-3)\\
L & (3L-4) & (2L-1)
\end{array}\right\} 
\end{equation}

Since the state $\Psi$ does not exist we must have A=0 and B=0.

But we can here actually prove that the state with $L_{T}=(3L-4)$
does not exist. The maximum M state for 3 L electrons with S=3/2 is
equal to L+(L-1) +(L-2) =3L-3. Thus we have a state with L$_{T}$
= L$_{max}$= ( 3L-3) M$_{T}$ = (3L-3). There is only one way to
form a state with M=3L-4. The M values of the 3 electrons are L, (L-1)
and (L-3). This state must be part of the (3L-3) multiplet. Thus,
we cannot have a state with S=3/2 L$_{T}$= (3L-4).

From what was mentioned above there also cannot be states with L$_{T}=$3L
and 3L-1 and 3L-2. Concerning the latter we can nevertheless try to
antisymmetrize the state $\Psi$= {[}{[}LL{]}$^{2L-1}$L{]}$^{L_{T}}$
where $L_{T}$ can be either (3L-2) or (3L-1). This leads to the following
relation 
\begin{equation}
1-2(4L-1)\left\{ \begin{array}{ccc}
L & L & (2L-1)\\
L & L_{T} & (2L-1)
\end{array}\right\} =0.
\end{equation}

As an example of the considerations in this section we see that for
3 electrons in the g shell we cannot have quartet states with $L_{T}$
equal to 2, 8, 10, 11, and 12.

\section{Bosons -- the odd L case}

We can make use of the vanishing 6-j of Eq. (2) for odd L. Consider
3 spinless L bosons with L odd with total angular momentum J=2. We
construct a symmetric wave function {[}{[}L(1) L(2){]}$^{J_{0}}$
L(3){]}$^{2}$ + {[}{[}L(3) L(2){]}$^{J_{0}}$ L(1){]}$^{2}$+{[}{[}L(1)
L(3){]}$^{J_{0}}$L(2){]}$^{2}$ Here J$_{0}$ can be L-1 or L+1.
Using Racah algebra this equals 
\begin{equation}
{[}1+2\Sigma{[}(2J_{0}+1)(2J_{a}+1){]}^{1/2}\left\{ \begin{array}{ccc}
L & L & J_{0}\\
L & 2 & J_{a}
\end{array}\right\} {]}{[}L(1)L(2){]}^{J_{a}}L(3){]}^{2}\qquad(J_{a}even)
\end{equation}

Taking J$_{0}$=L-1 we find

\begin{equation}
\left\{ \begin{array}{ccc}
L & L & L-1\\
L & 2 & L-1
\end{array}\right\} =-\frac{1}{2(2L-1)}
\end{equation}
 so the J$_{a}$=L-1 term vanishes but what about the J$_{a}$=L+1
term? We find that indeed

\begin{equation}
\left\{ \begin{array}{ccc}
L & L & L-1\\
L & 2 & L+1
\end{array}\right\} =0.
\end{equation}
 This is one of many cases of non-trivial vanishings of certain 6-j
symbols. Hence there will be no J=2 states of 3 bosons with odd L.
Note that this is the same condition as Eq. (2). For L=2 this condition
explains why there is no $^{4}$D state for 3 electrons (or identical
fermions e.g. neutrons) of the g$^{3}$ configuration. For L=3 it
explains why there are no J=2 states of 3 spinless bosons of the L$^{3}$
configuration.

It is also true that for bosons in a single L shell there is no state
with J=J$_{max}$-1. One can see this from the tables of Bayman and
Lande \cite{bayman66} and the online book of Dommelen \cite{vandono}.
One can also show it analytically.

The value of J$_{max}$ is nL where n is the number of bosons. This
is also the value of M$_{max.}$ One can construct a state with M=M$_{max}$-1
by changing the M value of the i'th particle to L-1. Call such a many-particle
state $\psi_{i}$. There are n such states but the only symmetric
wave function is $\sum\psi_{i}$. This state must belong to the J$_{max}$
multiplet and so there cannot be any state of n bosons in a single
L shell with J=J$_{max}$-1--same as in the fermion case.

One can nevertheless try to construct such a state by coupling two
L bosons to L$_{G}$ = 2L and symmetrizing the state {[}{[}LL{]}$^{2L}$L{]}$^{(3L-1)}$.
The non-existence of this state leads to a condition that is valid
not only for odd L but also for even L and half-integer L. 
\begin{equation}
1+4(4L+1)(-1)^{2L}\left\{ \begin{array}{ccc}
L & L & 2L\\
L & (3L-1) & 2L
\end{array}\right\} =0
\end{equation}

\section{Fermions in jj coupling}

The conditions in Eq.(2) and Eq. (3) which were originally derived
for even L, it not only holds for odd L but also for half integer
spin and is therefore useful in jj coupling situations. Indeed Talmi
\cite{talmi93} obtained this result by constructing a coefficient
of fractional parentage to a state of three neutrons in a single j
shell which he knew did not exist. In particular J$_{max}$ for 3
identical fermions is equal to $M_{max}=j+(j-1)+(j-2)=3j-3$. There
is only one state with M=M$_{max}$-1. One moves a nucleon with $M=(j-2)$
to the state with $M=(j-3)$. This state must belong to the J$_{max}$
multiplet, so there cannot be a state with $J=J_{max}-1$. Trying
to calculate cfp's to the nonexistent state leads to the result:

\begin{equation}
\left\{ \begin{array}{ccc}
j & (3j-4) & (2j-1)\\
j & j & (2j-3)
\end{array}\right\} =0.
\end{equation}
 This is the same as Eq. (3) but for half integer j greater than 3/2.

Sometimes the vanishing 6-j's are part of a bigger picture. Robinson
and Zamick \cite{dow1} used this relationship along with some 'diagonal
conditions' to demonstrate that for a system of two protons and one
neutron in a single j shell a partial dynamical symmetry (PDS) occurred
when one sets all two body matrix elements with T=0 to zero in a shell
model calculation. It turns out that not only J but also $J_{p}$
and $J_{n}$ separately are good quantum numbers. Furthermore states
with the same $J_{p}$ and $J_{n}$ are degenerate. The diagonal conditions
are

\begin{equation}
\left\{ \begin{array}{ccc}
j & j & (2j-1)\\
j & I & (2j-1)
\end{array}\right\} =\frac{(-1)^{2j}}{8j-2}
\end{equation}
 where I=(2j-1), (3j-2), and (3j-4).

Another known fact is that there are no J=$\frac{1}{2}$ states for
three identical fermions in a single j shell. This problem has been
addressed by Talmi \cite{talmi93,talmi05} and Zhao and Arima \cite{zhao03}.
We include this case here for completeness. To show the consequences
of this for 6-j relations we first couple two fermions to an even
angular momentum $J_{0}$. If j+1/2 is even, $J_{0}$ must be j+1/2;
if odd, $J_{0}$ must be j-1/2. We then add the third fermion and
couple the combination to $J=1/2$. We then antisymmetrize. The fact
that J=1/2 states for three fermions do not exist lead to the following
relations:

If j+1/2 is even we get 
\begin{equation}
1+2(2j+2)\left\{ \begin{array}{ccc}
j & j & (j+1/2)\\
j & 1/2 & (j+1/2)
\end{array}\right\} =0
\end{equation}

If j+1/2 is odd we get 
\begin{equation}
1+4j\left\{ \begin{array}{ccc}
j & j & (j-1/2)\\
j & 1/2 & (j-1/2)
\end{array}\right\} =0
\end{equation}

\section{Vanishing 9-j's}

From the fact that some states do not exist for four fermions in a
j shell Robinson and Zamick \cite{dow1,dow2} were able to show that
certain 9-j symbols vanished (see also related work by Zhao and Arima\cite{zhao05}).
This was an extension of the above arguments from Talmi about 3 fermions\cite{talmi93}.
One of their results is:

\begin{equation}
\left\{ \begin{array}{ccc}
j & j & (2j-1)\\
j & j & (2j-1)\\
(2j-1) & (2j-3) & (4j-4)
\end{array}\right\} =0
\end{equation}

They also used this relation for a different physical problem--a system
of 2 neutrons and 2 protons. If one sets all the two body interaction
matrix elements with isospin T=0 to zero then a partial dynamical
symmetry (PDS) emerges. There are certain angular momenta for this
system that cannot occur for a system of four identical fermions.

The PDS applies to these angular momenta. It turns out that not only
total J but also J$_{p}$ and J$_{n}$ separately are good quantum
numbers, and this is carried by the vanishing of the above 9-j symbol.
Furthermore states with the same J$_{p}$and J$_{n}$ are degenerate.

The diagonal conditions are 
\begin{equation}
\left\{ \begin{array}{ccc}
j & j & (2j-3)\\
j & j & (2j-1)\\
(2j-3) & (2j-1) & I
\end{array}\right\} =\frac{1}{4(4j-5)(4j-1)}
\end{equation}
 for I=(4j-4) (4j-5) and (4j-7) and 
\begin{equation}
\left\{ \begin{array}{ccc}
j & j & (2j-1)\\
j & j & (2j-1)\\
(2j-1) & (2j-1) & I
\end{array}\right\} =\frac{1}{2(4j-1)^{2}}
\end{equation}
 for I=(4j-4) (4j-2).

Let us now apply these results in more detail. For three identical
particles in a j shell the maximum J is $j+(j-1)+(j-2)=(3j-3)$. For
one proton and 2 neutrons the maximum value is $(2j-1)+j=(3j-1)$.
Hence states with J=3j-2 and 3j-1 are part of the PDS. These have
high spins and so the single j model might work better. Also belonging
to the PDS are states with J=$\frac{1}{2}$ and $J_{max}=3j-4$, the
last one belongs because there are no states with J=J$_{max}$ -1
for identical fermions (also true for identical bosons).

For 4 nucleons (or holes) the maximum J is $j+(j-1)+(j-2)+(j-3)=4j-6$.
However for two protons and 2 neutrons the maximum J is $(2j-1)+(2j-1)=4j-2$.
Hence states with J= (4j-5), (4j-4), (4j-3) and (4j-2) belong to the
PDS. These are again high spin states so the single j shell might
work fairly well for these. There might be other states with PDS e.g.
as noted above J=3 and 7 in the f$_{7/2}$ shell.

Consider next 3 nucleons in the g$_{9/2}$ shell. If they are identical
J$_{max}=\frac{21}{2}$. For a system of two protons and one neutron
the value of J$_{max}$ equals $\frac{25}{2}$. We get a degenerate
set J$_{p}$=9, J$_{n}=\frac{9}{2}$ with total angular momenta $J=\frac{19}{2}$,
$\frac{23}{2}$ and $\frac{25}{2}$ all with isospin T=$\frac{1}{2}$.

Consider four nucleons in the g$_{9/2}$ shell. If they are all identical,
J$_{max}$=12. For 2 protons 2 neutrons J$_{max}$=16. Here are selected
sets of degenerate states for four nucleons in the g$_{9/2}$ shell.
\\
 $\begin{array}{cccc}
J_{p} & J_{n}\\
8 & 8 & J=14,16 & T=0\\
8 & 6 & J=11,13,14 & T=0
\end{array}$

There are more. In the above (8,6) is an abbreviation for (8,6)+$(-1)^{J+T}$(6,8).
For the (8,6) configuration there is also a degeneracy of J=8 and
9. The above considerations do not explain this.

The discovery of the J=16$^{+}$ isomeric state in $^{96}$Cd has
been recently reported \cite{last}. It lies below the J=14$^{+}$
state and so the long lifetime is due to a spin gap. It can only decay
by gamma emission to a 10$^{+}$ state via an E6 transition and this
is highly inhibited. With only T=1 2 body matrix elements in the g9/2
shell the J=16$^{+}$ and J=14$^{+}$ would be degenerate as just
noted. Evidently the T=0 part of the 2-body interaction moves the
J=14$^{+}$ state above J=16$^{+}$.

\section{Vanishing 6-j in the f Shell-Racah, Judd and Regge}

Racah noted that for electrons in the f shell the calculation of coefficients
of fractional parentage could be greatly simplified by noting that
the exceptional group G2 is a subgroup of SO$_{7}$ \cite{racah49}.

The proof involved noting the following 6-j relation- $\left\{ \begin{array}{ccc}
5 & 5 & 3\\
3 & 3 & 3
\end{array}\right\} $ =0

Regge \cite{regge59} found several symmetry relations for 6-j symbols,
one of which is 
\begin{equation}
\left\{ \begin{array}{ccc}
a & b & e\\
d & c & f
\end{array}\right\} =\left\{ \begin{array}{ccc}
a & 1/2(b+c+e-f) & 1/2(b-c+e+f)\\
d & 1/2(b+c-e+f) & 1/2(-b+c+e+f)
\end{array}\right\} 
\end{equation}

Early on Judd \cite{judd70} used this to show that 
\begin{equation}
\left\{ \begin{array}{ccc}
5 & 5 & 3\\
3 & 3 & 3
\end{array}\right\} =\left\{ \begin{array}{ccc}
5 & 4 & 4\\
3 & 4 & 2
\end{array}\right\} 
\end{equation}

See also the work of Judd and Li\cite{judd89}. Furthermore we emphasized
at the beginning of this work that for quartet states of three electrons
in the g shell shell the space wave function has to be antisymmetric.
This leads to the vanishing of the 6-j on the right hand side above.
This is easier to understand than the f shell result of Racah \cite{racah49}.
Thus we have an amusing connection between electrons in the f shell
and those in the g shell and some of the mystery of the vanishing
Racah has been removed. The above result has also been used by G Vanden
Berghe et al. \cite{berghe84}. They show many other examples of vanishing
6-j's.

It should be noted that the following Regge symmetry relation,(previously
used by Robinson and Zamick \cite{dow1}) can shed some light on the
relation between Eq (2) and Eq.(3): 
\begin{equation}
\left\{ \begin{array}{ccc}
a & b & c\\
d & e & f
\end{array}\right\} =\left\{ \begin{array}{ccc}
(b+c+f-e)/2 & (a+c+d-f)/2 & (a+b+e-d)/2\\
(c+e+f-b)/2 & (a+d+f-e)/2 & (d+c+b-a)/2
\end{array}\right\} 
\end{equation}

This leads to the relation 
\begin{equation}
\left\{ \begin{array}{ccc}
j & j & (2j-3\\
(3j-4) & j & (2j-1)
\end{array}\right\} =\left\{ \begin{array}{ccc}
(2j-2) & (2j-3) & 2\\
(2j-2) & (2j-1) & (2j-2)
\end{array}\right\} 
\end{equation}

This is true for even j, odd j and half-integer j. We have already
shown that the 6-j on the left which appears in Eq. 3 vanishes for
even j and half-integer j by constructing all m states for M$_{max}$=
J$_{max}$ and J$_{max}$ -1. This did not involve 6-j symbols explicitly.
We can now use this relation to then show selected vanishings in Eq.
2 e.g.

$\left\{ \begin{array}{ccc}
4 & 4 & 5\\
8 & 4 & 7
\end{array}\right\} $= $\left\{ \begin{array}{ccc}
6 & 2 & 5\\
6 & 6 & 7
\end{array}\right\} $=0 $\qquad\qquad\qquad$ $\qquad\qquad\qquad\left\{ \begin{array}{ccc}
7/2 & 7/2 & 4\\
13/2 & 7/2 & 6
\end{array}\right\} $=$\left\{ \begin{array}{ccc}
5 & 4 & 2\\
5 & 6 & 5
\end{array}\right\} $ =0

We thus estabish connections between 6-j's whose vanishing can be
obtained from m state arguments to selected states which have {}``two''
in them as per Eq. 2.

We note explicit expressions for that 6j symbols with a \textquotedblleft{}two\textquotedblright{}
in them have been worked out by Biedenharn, Blatt, and Rose \cite{bbr}.
Using their notation we fi{}nd from their results that $\left\{ \begin{array}{ccc}
l_{1} & J_{1} & 2\\
J_{2} & l_{2} & L
\end{array}\right\} $ for $l_{2}=J_{1}+1$ and $l_{1}=J_{1}+1$ is proportional to X where
$X=[(J_{1}+1)(J_{1}\lyxmathsym{\textminus}J_{2})\lyxmathsym{\textminus}L(L+1)+J_{2}(J_{2}+2)]$.
We have L = 2j \textminus{} 2, J$_{1}$ = 2j \textminus{} 3, l$_{1}$
= 2j \textminus{} 2, J$_{2}$ = 2j \textminus{} 2, and l$_{2}$= 2j
\textminus{} 1. With these values we see that X vanishes.

\section{Closing}

I$ $n summary we have in this work mainly addressed the problem of
missing states for fermions in LS coupling, especially electrons,
fermions in jj coupling, especially for particles of one kind e.g.
neutrons only or protons only, and of bosons. We note that very similar
expressions apply in the different cases. For example the non-existence
of quartet (S=3/2) states with total angular orbital momentum L$_{T}$=
2 for an L$^{3}$ configuration with even L is closely associated
with the non-existence of spinless boson states also with L$_{T}$=2
but for an odd L, L$^{3}$ configuration. We also have shown that
in all three cases states with J=J$_{max}$ -1 did not exist. We were
able to obtain these results not only in terms of vanishing 6-j symbols
but also by counting the number of m states. On the other side we
have shown that the value of having 6-j symmetries is greatly enhanced
by the Regge symmetry relations . They help to establish connections
with what were a priori diverse subjects. By putting all these results
in one place we hope we have conveyed the beautiful unity that pervades
the problem of missing states.

This work can be regarded as an extension of previous work on companion
problems \cite{zamick05}. In the previous contributions we showed
how similar expressions have consequences on different physical problems
and different branches of physics e.g. how isospin can be used to
get the same results as quasispin \cite{zamick05,steele03}. We find
such such associations fascinating. In this work we show that an expression
involving even L in which was used to explain the absence of certain
states in L-S coupling can be generalized to odd L in order to explain
the absence of certain bosonic states and can also be generalized
to half integer angular momenta to explain the absence of certain
states in jj coupling,the latter being most relevant to nuclear physics.

Acknowledgments:

Part of this work was performed at the Weizmann Institute where L.Z.
was a Morris Belkin visiting professor. L.Z. thanks Igal Talmi for
his suggestions, guidance and hospitality. L.Z. thanks Brian Judd
for making him aware of his work. S.J.Q.R. thanks Millsaps College
for a faculty development grant to support this work.


\begin{thebibliography}{References}
\bibitem{judd89} J.B. Judd and Shaozhong Li, J. Phys. B: Atm. Mol.
Opt. Phys. 22 (1989) 2057-2070

\bibitem{talmi93} I.Talmi Simple Models of Complex Nuclei,Harwood
academic publications 1993

\bibitem{dow1} S.J.Q. Robinson and L.Zamick, Phys. Rev. C63,064316
(2001)

\bibitem{bayman66} B.F. Bayman and A. Lande, Nucl. Phys. 77 (1966)
1

\bibitem{vandono} L. Van Dommelen,Angular Momentum for Engineers,
5.26 alpha, online book .http://www.eng.fsu.edu/~dommelen/quantum/

\bibitem{talmi05} I. Talmi, Phys. Rev. C72, 037302 (2005)

\bibitem{zhao03} M. Zhao and A. Arima, Phys. Rev. C68, 044310 (2003)

\bibitem{dow2} S.J.Q. Robinson and L. Zamick, Phys. Rev. C64,057302
(2001)

\bibitem{zhao05} Y.M. Zhao and A. Arima Phys. Rev. C72, 054307 (2005)

\bibitem{zamick05} L. Zamick and A. Escuderos, Phys. Rev. C71, 014315
(2005)

\bibitem{racah49} G. Racah, Phys. Rev. 76 , 1352 (1949)

\bibitem{regge59} T. Regge, Il Nuovo Cimento Vol XI, N1 298 (1959)

\bibitem{last} B.S. Nara Singh, N-Z physics near A-100, Post conference
workshop Nuclear physics in Astrophysics Weizmann Institute, Isreal,
April 10-11 (2011) online at http://www.weizmann.ac.il/conferences/NPA5/workshop

\bibitem{judd70} B.R. Judd and J.P. Elliott,Topics in Atomic and
Nuclear Theory, The Caxton Press Christchurch (1970)

\bibitem{berghe84} G. Vanden Berghe , H.D. De Meyer, and J. Van der
Jeugt, J. Math. Phys. 25 ,2585 (1984)

\bibitem{steele03} G. Rosensteel and D.J. Rowe, Phys. Rev. C67, 014303
(2003)

\bibitem{bbr} L.C. Biedenharn, J.M. Blatt, and M.E. Rose, Review
of Modern Physics, 24 212 (1952)\end{thebibliography}
\end{document}